\documentstyle[preprint,prd,eqsecnum,aps]{revtex}
\tighten
\begin{document}
\draft

\title {On the variational principle for dust shells in General Relativity}
\author{Valentin D. Gladush\footnote{E-mail: gladush@ff.dsu.dp.ua}}
\address{Department of Physics, Dnepropetrovsk State University, \\
per. Nauchniy 13, Dnepropetrovsk 49050, Ukraine}
\date{\today}
\maketitle

\begin{abstract} The variational principle for a thin dust shell in General
Relativity is constructed. The principle is compatible with the
boundary-value problem of the corresponding Euler-Lagrange equations, and
leads to ``natural boundary conditions'' on the shell. These conditions
and the gravitational field equations which follow from an initial
variational principle, are used for elimination of the gravitational
degrees of  freedom. The transformation of the variational formula for
spherically-symmetric systems leads to two natural variants of the
effective action.  One of these variants describes the shell from a
stationary interior observer's point of view, another from the exterior
one. The conditions of isometry of the exterior and interior faces of the
shell lead to the momentum and Hamiltonian constraints. The canonical
equivalence of the mentioned systems is shown in the extended phase space.
Some particular cases are considered.
\end{abstract}

 \narrowtext
\newpage
\section{INTRODUCTION}

A thin spherically-symmetric dust shell is among the simplest popular
models of collapsing gravitating configurations. The equations of motion
of these objects are obtained in \cite{israel1}, \cite{kuchar1}. The
construction of a variational principle for such systems was discussed
from different points of view in \cite{visser}-\cite{berezin3}. There are
a number of problems here, most basic of which is the dependence on the
choice of the evolution parameter (internal, external, proper). The choice
of time coordinate, in turn, affects the choice of a particular
quantization scheme, leading, in general, to quantum theories which are
not unitarily equivalent.

In most of these papers the variational principle for shells is usually
constructed in a comoving frame of reference, or in one of variants of
freely falling frames of reference. However, use of such frames of
reference frequently leads to effects unrelated to the object under
consideration. The essential physics involves a picture of a gravitational
collapse from the point of view of an infinitely remote stationary
observer. In quantum theory this point of view enables us to treat bound
states in terms of asymptotic quantities and to build the relevant
scattering theory correctly. On the other hand, to treat primordial black
holes in the theory of self-gravitating shells it is convenient to take
the viewpoint of a central stationary observer. In the approach related to
proper time of the shell reduction of the system leads to complicated
Lagrangians and Hamiltonians which creates difficulties on quantization.
In particular it leads to theories with higher derivatives or to finite
difference equations.

In our opinion, the choice of the exterior or interior stationary
observers is most natural and corresponds to the real physics. To provide
the necessary properties of invariance, specification of the canonical
transformations in an extended  phase space which translate the
corresponding dynamical systems into one another enough. In addition the
action for a shell should satisfy some natural requirements. In the
absence of self-forces it should pass into the action for a geodesic
motion. Further, according to the correspondence principle, at small
velocities and masses of the shell, and also in the absence of other
sources of the gravitational field, we should obtain the action for a
self-gravitating Newtonian shell (see Appendix C).

The natural Hamiltonian formulation of a self-gravitating shell was
considered in works \cite{hajicek5}, \cite{dolgov}. However this
formulation was not obtained by a variational procedure from some initial
action containing the standard Einstein-Hilbert  term. The action for such
a self-gravitating spherical shell of mass $m$ can be introduced with the
help of a naive ``relativization'' of the Newtonian action. It is carried
out by simple replacement of the kinetic energy $mv^2/2$ by the
relativistic expression $-mc\sqrt{(1-v^2/c^2)}$ (see below Lagrangians
(\ref{2.7}) and (\ref{4.2.23k})). If there is an exterior gravitational
field then the kinetic and potential energy of the shell have as their
general relativistic analog the geodesic Lagrangian
$-mc~^{(2)}ds_{\pm}/dt_{\pm}$. The subsigns ``$_{\pm}$'' correspond to
exterior and interior observers. The gravitational self-action of the
shell is the same for all cases, and its sign depends on whether
stationary observer can exist inside and outside the shell.

The above  Hamiltonian  formulation for the shell, as well as the
procedure of ``relativization'' follows from the Lagrange formalism of
dust shells constructed in the present paper. We view the system as a
compound configuration consisting of two vacuum regions with a
spatially-closed boundary surface formed by the shell. The initial action
we take as the sum of actions of York type for either region and the
action for dust matter. For the complete action introduced in this way the
variational principle is compatible with the boundary-value problem of the
corresponding Euler-Lagrange equations for either region of the
configuration, and leads to ``natural boundary conditions'' on the shell.
The missing boundary conditions are obtained by consideration of the
variations with respect to normal displacements of the shell. The obtained
conditions coincide with the known Israel matching conditions at singular
hypersurfaces and are considered as constraints. Together with the
equations of the gravitational field they are used to eliminate of the
gravitational degrees of freedom. The tangential variations of
thus-obtained action with constraints lead to the known equations of
motion of the Israel \cite{israel1}.

The problem of the complete reduction of the action is solved for
spherically-symmetric systems. By transforming the variational formula and
using the constraints the obtained action is reduced to two variants of
the effective action. One of these variants describes the shell from an
interior stationary observer's point of view, and the other from the
exterior one. Then we go over from the Lagrangian to the Hamiltonian
description. The conditions of isometry of the exterior and interior sides
of the shell lead to the momentum and Hamiltonian constraints. The
canonical equivalence of these two variants of the description of the
shells in the extended phase space indicates the existence of a ``discrete
gauge'' transformation associated with the transition from the interior
observer to the exterior one.

The paper is organized  as follows. In Sec.II the full action is
constructed for a compound, piecewise smooth Lorentz manifold with a
four-dimensional spatially-closed boundary surface between two vacuum
regions, corresponding to the world sheet of the shell. From here the
Einstein equations for regions outside the shell and surface equations
follow. Further, the action for the shell and the equations of motion are
constructed.

In Sec.III spherically-symmetric relativistic dust shells are considered.
The Lagrangians and Hamiltonians describing the shell from the point of
view of the interior or exterior observer are obtained. Then momentum and
Hamiltonian constraints are found. They emerge from independent
consideration of the interior and exterior faces of the shell using the
conditions of isometry of its two faces. In Sec.IY special cases of dust
shells and configurations of several shells are briefly considered.

In Appendix A it is shown that the surface equations, obtained in Sec.II,
reduce to the known equations for jumps of the extrinsic curvature tensor
of the shell. In Appendix B we show the canonical equivalence of the
actions for the dust spherically-symmetric shell written relative to the
interior and exterior observers. This equivalence is thought of as
operating in the extended phase space of the corresponding dynamical
system. In Appendix C the action for an arbitrary nonrelativistic
gravitating dust shell is constructed.  The Lagrangian for the spherical
gravitating nonrelativistic dust shell is found. It was deemed worthwhile
to consider the nonrelativistic case because it clarifies the
interpretation of the results and allows comparisons with the general
relativistic approach.

In this work we consider both relativistic and non-relativistic systems.
In this connection, we shall keep all the dimensional constants. Here $c$
is the velocity of light, $\gamma$ is the gravitational constant, $\chi =
8\pi\gamma/c^2$, $\hbar$ is Planck's constant. The metric tensor
$g_{\mu\nu} \ (\mu, \nu = 0,1,2,3)$ has the signature $(+\ -\ -\ -)$.

\section {The variational principle and equations of motion for
relativistic dust shells}

Consider a time-like spatially-closed hypersurface $\Sigma^{(3)}_t$ into
some region $D^{(4)}$ of the space-time $V^{(4)}$. Let it be the world
sheet of the infinitely thin dust shell with the surface density of dust
$\sigma$. This shell divides the region $D^{(4)}$ into the interior and
exterior ones, $D^{(4)}_-$ and $D^{(4)}_+$. Introduce the general
coordinate map $x^{\mu}$ on our compound manifold $D^{(4)}=D^{(4)}_{-}
\cup \Sigma^{(3)}_{t} \cup D^{(4)}_{+}$ and the metrics $g^{\pm}_{\mu\nu}$
on $D^{(4)}_{\pm}$, so that $g^{-}_{\mu\nu}|_{\Sigma^{(3)}_{t}} =
g^{-}_{\mu\nu}|_{\Sigma^{(3)}_{t}}$.

One defines the elements of the four-volume $d^4\Omega$ on $D^{(4)}_{\pm}$
and three-volume $d^3\Omega$ on $\Sigma^{(3)}_t$ according to the formulas
\begin{eqnarray}
      d^4\Omega & = & \sqrt{-g}d^4 x =
 \sqrt{-g}dx^0 \wedge dx^1 \wedge dx^2 \wedge dx^3 \, , \label{3.2} \\
      d^3\Omega & = & - \sqrt{-g} n^{\mu} d \Sigma_{\mu}
      = \sqrt{-g} d \Sigma \, ,                               \label{3.3}
\end{eqnarray}
where $n^{\mu}$ is the unit normal to $\Sigma^{(3)}_t$, directed from
$D^{(4)}_-$ to $D^{(4)}_+$ \ $(n_{\mu}n^{\mu}=-1,\, u_{\mu}n^{\mu}=0)$,
$g=\det|g_{\mu\nu}|$. Three-forms $d\Sigma_{\mu}$ and $d\Sigma$ are
determined by the relations
\begin{equation}
dx^{\mu} \wedge d\Sigma_{\nu} = \delta^{\mu}_{\nu}d^4 x \, , \qquad
 \eta \wedge d\Sigma = d^4 x \qquad (d\Sigma_{\mu} =n_{\mu}d\Sigma)\, ,
\end{equation}
where  ``$\wedge$'' denotes the exterior product, and
$\eta=n_{\mu}dx^{\mu}$ is a normal covector.

Now let us fix coordinate system $x^{\mu} $ so that the coordinates $x^a\
(a=2,3) $ be Lagrange coordinates of particles on the shell $
\Sigma^{(3)}_t$. Then $u^{\mu}x^a_{,\, \mu}=n^{\mu}x^a_{,\, \mu}=0$, where
``$_{,\, \mu}$'' is derivative with respect to the coordinate $x^{\mu}$.
Hence it follows $u^a=n^a=0$. The equations $x^a =\mbox{const}$ determine
the world line $\gamma$ of some particle of dust on $\Sigma^{(3)}_t$. The
set $ \{\gamma \} = \{ x^a,~x^a + dx^a\}$ of the world lines forms the
elementary stream tube of dust. On the shell $\Sigma^{(3)}_t$ we shall
introduce the basis of one-forms
\begin{equation}
      \{ e^0 \equiv \omega = u_{\mu}dx^{\mu}\, , \quad
      e^a = dx^a \} \qquad (a,b=2,3)  \label{3.4}
\end{equation}
and the dual vector basis
\begin{eqnarray}
      \{ e_0 \equiv u = u^{\mu} \partial _{\mu}\, , \quad e_a \}\, ,
      \qquad
      e^i (e_k) = \delta ^{i}_{k} \quad  (i,k = 0,2,3)\, . \label{3.5}
\end{eqnarray}
In the basis $\{e^i \}=\{\omega ,\  dx^a \}$ the metric tensor and
three-form of volume on $\Sigma^{(3)}_t$ are
\begin{eqnarray}
      ~^{(3)}g & = & \omega \otimes \omega - q _{ab}dx^a \otimes dx^b\, ,
       \label{3.6}  \\
      d^3\Omega = & - & \omega \wedge d^2\Omega \, ,   \qquad
      d^2\Omega = \sqrt{q} dx^2 \wedge dx^3 \, .   \label{3.8}
\end{eqnarray}
Here ``$\otimes $'' is the sign of a tensor product, $d^2\Omega$ is the
surface element of the area for the section which is orthogonal to the
elementary stream tube of dust, $q_{ab}$  is the metric on these sections,
$q =\det |q_{ab}|$. In the neighbourhood of the hypersurface
$\Sigma^{(3)}_t$ the metric tensor $~^{(4)}g$ and four-form of volume $d^4
\Omega $ can be expressed in the form
\begin{eqnarray}\label{3.7}
     ^{(4)}g & = & ^{(3)}g - \eta \otimes \eta \, , \\
     d^4\Omega & = & \eta \wedge  d^3\Omega
      = \omega \wedge \eta \wedge d^2\Omega \, .
\end{eqnarray}
We introduce the two-form of mass on $\Sigma^{(3)}_t$ by the formula $ d^2
m=\sigma d^2\Omega $,  then $\sigma d^3\Omega =\omega\wedge d^2 m$.

Now we take the full action of the compound configuration in the form
\begin{equation}\label{4.1}
     I^{(g)}_{tot} = I_{EH} - c\int\limits_{\Sigma^{(3)}_t}
     \left(\sigma n^{\mu} + \frac{1}{2\chi}[\omega^{\mu}]\right)
      \sqrt{-g} d \Sigma_{\mu} + I_{\partial D^{(4)}} + I_{0}\, .
\end{equation}
It is the functional of the metric $g_{\mu\nu} $, density of the dust $
\sigma $ and hypersurface $ \Sigma^{(3)}_t$: $I^{(g)}_{tot}\equiv
I^{(g)}_{tot} (g_{\mu\nu}, \ \sigma, \ \Sigma^{(3)}_t)$. The first term in
the right side of (\ref {4.1})
\begin{equation}\label{4.2}
      I_{EH} = - \frac{c}{2\chi} \int\limits_{D^{(4)}_- \cup D^{(4)}_+}
      ~^{(4)}R~d^4\Omega \,
\end{equation}
is the Einstein-Hilbert action for the regions $D^{(4)}_{\pm}$, where
$~^{(4)}R~$ is the curvature scalar.

The second term in the right side (\ref {4.1}) contains the matter term $
c\sigma d^3\Omega $ and matching term. The symbol
$[\omega^{\mu}]=\omega^{\mu}|_{+}-\omega^{\mu}|_{-}$ denotes the jump of
the quantity
\begin{eqnarray}
      \omega^{\mu} &=& g^{\sigma \rho} \Gamma^{\mu}_{\sigma \rho} -
                       g^{\mu \rho} \Gamma^{\sigma}_{\sigma \rho} \, ,
                           \label{4.4}  \\
       \Gamma^{\mu}_{\sigma \rho} &=& \frac{1}{2}g^{\mu \nu}
    (g_{\nu\rho ,\sigma}+g_{\nu\sigma ,\rho}-g_{\rho\sigma ,\nu}) \, ,
                        \label{4.5}
\end{eqnarray}
on $\Sigma_t $. The sign ``$|_+$'' or ``$|_-$'' indicates the marked
values to be calculated as a limiting magnitude when approaching the
boundary  $\Sigma_t$ from outside or inside respectively. In Appendix~A it
will be shown, that the relation
\begin{equation}\label{4.5a}
            [\omega^{\mu}]n_{\mu} =2[K]
\end{equation}
takes place. Here $K=g^{\mu\nu} K_{\mu\nu}$ is the trace of the extrinsic
curvature tensor
\begin{equation}\label{4.16a}
   K_{\mu\nu} = - n_{\mu;\rho}h^{\rho}_{\nu}\, \qquad
   (h^{\rho}_{\nu} = \delta ^{\rho}_{\nu} + n^{\rho}n_{\nu})\, ,
\end{equation}
where $_{;\,\rho}$ is covariant derivative with respect to the coordinate
$x^{\mu}$. The third term
\begin{equation}\label{4.3}
      I_{\partial D^{(4)}} = \frac{c}{2\chi} \int\limits_{\partial D^{(4)}}
             \omega^{\mu}\sqrt{-g} d \Sigma_{\mu}
\end{equation}
contains the surface terms which are introduced to fix the metric on the
boundary $\partial D^{(4)}$ of the region $ D^{(4)}$. Note, that the
boundary $\partial D^{(4)}$  consists of the pieces of time-like as well
as space-like hypersurfaces. The last term $I_{0}$ in (\ref{4.1}) contains
the boundary terms on the time-like infinitely remote hypersurfaces,
necessary for normalization of the action.

The relation
\begin{equation}\label{4.6}
      \sqrt{-g}~^{(4)}R =\sqrt{-g}G+(\sqrt{-g}\omega^{\mu})_{, \,\mu}
\end{equation}
takes place, where
\begin{equation}\label{4.7}
      G = g^{\mu \nu}\left(\Gamma^{\rho}_{\mu \sigma}\Gamma^{\sigma}_{\nu \rho}
          - \Gamma^{\sigma}_{\mu\nu} \Gamma^{\rho}_{\rho \sigma}\right)
\end{equation}
contains only the first derivatives of the metric. Therefore the action
(\ref{4.1}) can be rewritten in a more compact form
\begin{equation}\label{4.8}
      I^{(g)}_{tot} = I_{g} + I_{m} + I_{0}\, ,
\end{equation}
where
\begin{equation}\label{4.9}
      I_{g} = - \frac{c}{2\chi} \int\limits_{D^{(4)}_- \cup
      D^{(4)}_+}\!\! \sqrt{-g} G d^4 x =
              \int\limits_{D^{(4)}_- \cup D^{(4)}_+} L_{g} d^4 x
\end{equation}
is the gravitational action of the first order, and
\begin{equation}\label{4.10}
    I_{m} =  c\int\limits_{\Sigma^{(3)}_{t}}\sigma d^3\Omega =
   - c\int\limits_{S^{(2)}_{t}} d^2 m \int\limits_{\gamma} \omega
\end{equation}
is the action for the dust.

The first and the penultimate terms in (\ref{4.1}) form the action which
can be ascribed to that of the York's type $I_{Y}= I_{EH} + I_{\partial
D^{(4)}}$. It is used in variational problems with the fixed metric on the
boundary $\partial D^{(4)}$ of the region $ D^{(4)}$. It can also be used
in variational problems with the general relativistic version of ``natural
boundary conditions'' for ``free edge'' \cite{hayward1}. In this case the
metric on the boundary is arbitrary and the corresponding momenta
vanishes. Together with $I_{0}$ it forms the York-Gibbons-Hawking action $
I_{YGH}=I_{Y}+I_{0}$ for a free gravitational field.

In our case of the compound configuration we also fix the metric on
boundary $\partial D^{(4)}$, as it was done in variational problem for
action $I_{Y} $. In addition, inside the system there is the boundary
surface $\Sigma^{(3)}_t$, with singular distribution of matter on it. One
can interpret this configuration as the two vacuum regions $D^{(4)}_{\pm}$
with a common ``loaded edge'' (or with a ``massive edge''). The sum of the
actions of type $I_{Y}$ for these regions and of the action for matter
$I_{m} $ and normalizing term $I_{0} $ do leads to the action
$I^{(g)}_{tot}$.

If there is no dust, $\sigma =0$, the common boundary is not ``loaded''.
Then, the requirement $\delta I^{(g)}_{tot}=0 $, at arbitrary, everywhere
continuous variations of the metric, gives generalization of the above
``natural boundary conditions'' for free hypersurface $ \Sigma^{(3)}_t$.
They coincide with the condition of continuity for the extrinsic curvature
on $ \Sigma^{(3)}_t$, i.e., with ordinary matching conditions. If the
edge, being matched, is ``loaded'' by some surface distribution of matter,
then we obtain the corresponding surface equation or the boundary
conditions for $D^{(4)}_{\pm} $. They are the analog of the generalized
``natural boundary conditions'' for ``loaded edges''. The initial action
is chosen so, that the surface equations on $\Sigma^{(3)}_t$ following
from the requirement $\delta I^{(g)}_{tot}=0$, coincide with the matching
conditions on singular hypersurfaces \cite{israel1}. In this case, the
variational principle for the action $I^{(g)}_{tot}$ will be compatible
with the boundary-value problem of the corresponding Euler-Lagrange
equations \cite{kurant}, \cite{ponomarev}.

Note, that, as a rule, the boundary terms are formulated in terms of the
extrinsic curvature of the corresponding hypersurfaces. For the
configuration which contains the boundary hypersurface dividing the domain
$D^{(4)}$ into parts and the whole boundary consisting of several pieces
of edge, initial, and eventual hypersurfaces, it is more convenient to use
the covariant approach. In order to calculate $\delta I^{(g)}_{tot}$ we
use the complete action in the form (\ref{4.1}). According to
\cite{landau2} we have
\begin{equation}\label{4.11}
   \delta \left(\sqrt{-g}~^{(4)}R\right)
   =-\sqrt{-g}~^{(4)} G^{\mu\nu} \delta g_{\mu\nu}
   + \left(\sqrt{-g} \Omega ^{\mu}\right)_{,\, \mu}\, ,
\end{equation}
where
\begin{equation}\label{4.12}
       \Omega ^{\mu} =  g^{\sigma \rho} \delta \Gamma^{\mu}_{\sigma \rho} -
              g^{\mu \rho} \delta \Gamma^{\sigma}_{\sigma \rho}\, ,
\end{equation}
and $~^{(4)}G^{\mu\nu} =~^{(4)}R^{\mu\nu}-\frac{1}{2}~^{(4)}R g^{\mu\nu}$
is the Einstein tensor. In addition, we shall use the following
conditions: the boundary of the configuration $\partial D^{(4)}$, the
metric on it, and the normal vector are fixed. Then $\delta d\Sigma_{\mu}
|_{\partial D^{(4)}} =0,\  \delta g_{\mu\nu}|_{\partial D^{(4)}} =0,\
\delta n_{\mu}|_{\partial D^{(4)}}=0 $. The hypersurface $\Sigma^{(3)}_t$
is fixed, and the metric and its variations are continuous on
$\Sigma^{(3)}_t$: $~[g_{\mu\nu}]_{\Sigma^{(3)}_t} =0 $, $[~\delta
g_{\mu\nu}]_{\Sigma^{(3)}_t} = 0 $, $ [n_{\mu}]_{\Sigma^{(3)}_t} =0,\
[\delta n_{\mu}]_{\Sigma^{(3)}_t} =0 $.

For the variation $\delta I_{m}$ according to the formula (\ref{4.10}) we
have $\delta I_{m}=- c\int d^2 m \int\delta\omega =- c\int d^2 m \int
\delta\omega_{\gamma}$. Here, the quantity $d^2m$ is considered as a
stationary value at variations of the metric \cite{haw}.  The sign
``$~_{|\gamma}$'' designates restriction of the one-forms on the world
line $\gamma $ so, that
\begin{equation}
  \delta\omega_{|\gamma}=\delta ds
  =\frac{1}{2}u^{\mu}u^{\nu}ds~\delta g_{\mu\nu}
  =- \frac{1}{2}u_{\mu}u_{\nu} \omega_{|\gamma}~\delta g^{\mu\nu}\, .
\end{equation}

If all these conditions are satisfied, then from the requirement $\delta
I^{(g)}_{tot} =0$ one obtains the vacuum Einstein equations
\begin{equation}\label{4.13}
   ~^{(4)} G^{\mu\nu} = 0\, , \qquad \forall D^{(4)}_{\pm}
\end{equation}
and the surface equations on $ \Sigma^{(3)}_{t}$
\begin{equation}\label{4.14}
      Q_{\mu\nu} - \frac{1}{2}~Q g_{\mu\nu} = -\chi \sigma u_{\mu}u_{\nu}\, ,
\end{equation}
where $Q=g^{\mu\nu}Q_{\mu\nu}$, and
\begin{equation}\label{4.15}
      Q_{\sigma \rho} = n_{\mu}[\Gamma^{\mu}_{\sigma \rho}]
                 -\frac{1}{2}\left(n_{\sigma}[\Gamma^{\mu}_{\mu \rho}]
                  + n_{\rho}[\Gamma^{\mu}_{\mu \sigma}]\right).
\end{equation}
It is shown in Appendix A that the surface equations (\ref{4.14}) reduce
to the known equations for the jump discontinuity of the extrinsic
curvature tensor of the hypersurface  $\Sigma^{(3)}_t$ \cite{israel1}
\begin{eqnarray}\label{4.16}
     [K_{\mu\nu}] - [K] h_{\mu\nu}
      =- \chi \sigma u_{\mu}u_{\nu}\, ,
\end{eqnarray}
where $ h_{\mu\nu} = g_{\mu\nu}+n_{\mu}n_{\nu} $ is the metric on $
\Sigma^{(3)}_t $. From the relations (\ref{4.16}) it follows
\begin{equation}\label{4.16b}
  [ K_{\mu \nu }] u^{\mu}u^{\nu} =- \frac{\chi}{2} \sigma \, ,
\end{equation}

The missing equation for the average tensor of the extrinsic curvature
\begin{equation}\label{middle-K}
  \bar K^{\mu}_{\nu}=\frac{1}{2}(K^{\mu}_{\nu {|_{+}}}+K^{\mu}_{\nu
{|_{-}}})
\end{equation}
can be obtained by considering the variations of $I^{(g)}_{tot}$ with
respect to normal displacements of the hypersurface $\Sigma^{(3)}_t$. For
this purpose we define some one-parameter family of time-like
hypersurfaces in a neighbourhood of $\Sigma^{(3)}_t$ so that
$\Sigma^{(3)}_t$ is included in this family. The family induces
(3+1)-decomposition of
 the objects in the neighbourhood of
$\Sigma^{(3)}_t$. Thus for the four-curvature scalar one has
\begin{equation}\label{4.17}
      ~^{(4)} R =~^{(3)} R + K^{\mu}_{\nu}K^{\nu}_{\mu} - K^2
  +\frac{2}{\sqrt{-g}}\left\{\sqrt{-g} (Kn^{\mu}-a^{\mu})\right\}_{,\mu}\, ,
\end{equation}
where $a^{\mu}=n^{\mu}_{;\,\nu}n^{\nu}$ and $~^{(3)} R$ is the curvature
scalar of hypersurfaces of the family. Substituting (\ref{4.17}) into
(\ref{4.2}) and taking into account the relations $a^{\mu}n_{\mu}=0$ and
(\ref{4.5a}), one obtains the action (\ref{4.1}) in the form
\begin{equation}\label{4.18}
   I^{(g)}_{tot} =\hat I_{g} + I_{m} + \hat I_{\partial D^{(4)}} + I_{0}\, ,
\end{equation}
where
\begin{equation}\label{4.19}
    \hat I_{g} = \int\limits_{D^{(4)}_- \cup D^{(4)}_+}\!\!\hat L_{g} d^4 x
  = - \frac{c}{2\chi} \int\limits_{D^{(4)}_- \cup D^{(4)}_+}\!\!
  \left(^{(3)}R+K^{\mu}_{\nu}K^{\nu}_{\mu}-K^2 \right)\sqrt{-g}~d^4 x \,
\end{equation}
is the gravitational action, containing normal derivatives up to the first
order, and $\hat I_{\partial D^{(4)}} $ and $I_{0}$ contain the boundary
terms, which are unessential here.

Now let every point $p\in\Sigma^{(3)}_t $ be translated at a coordinate
distance $\delta x^{\mu}(p)=n^{\mu}\delta\lambda (p)$ in the normal
direction. As a result of the displacement one gets a new hypersurface
${\tilde \Sigma}^{(3)}_t $. The initial and eventual positions of a shell
are fixed, therefore $\delta\lambda (p)=0, \ \forall p\in\Sigma^{(3)}_t
\cap\partial D^{(4)}={\tilde\Sigma}^{(3)}_t \cap\partial D^{(4)}$. In
addition, we fix the metric $g_{\mu\nu}$ and all the quantities on $
\Sigma ^{(3)}_t $, so that $\delta I_{m}=0$.

As a result of the displacement of the hypersurface $\Sigma^{(3)}_t $, the
initial regions $D^{(4)}_{+}$ and $D^{(4)}_{-}$ are transformed into new
ones ${\tilde D}^{(4)}_{+} $ and ${\tilde D}^{(4)}_{-}$, so that, ${\tilde
D}^{(4)}_{-} \cup {\tilde\Sigma}^{(3)}_{t} \cup {\tilde D}^{(4)}_{+} =
D^{(4)}_{-} \cup \Sigma^{(3)}_{t} \cup D^{(4)}_{+} = D^{(4)}$. Then, for
example, the variation of the region $ D^{(4)}_{-} $ can be expressed in
the form $\delta D^{(4)}_{-} = {\tilde D}^{(4)}_{-} \backslash D^{(4)}_{-}
= D^{(4)}_{+} \backslash {\tilde D}^{(4)}_{+}$. The variation of the
action (\ref{4.19}), under the above conditions, proves to be equal
\begin{equation}\label{4.20}
   \delta I^{(g)}_{tot} = \delta\hat I_{g} =
\int\limits_{{\tilde  D}^{(4)}_- \cup {\tilde D}^{(4)}_+}\!\!\hat L_{g}\
d^4 x
  - \int\limits_{D^{(4)}_- \cup D^{(4)}_+}\!\!\hat L_{g}\ d^4 x
  \cong - \int\limits_{\delta D^{(4)}_{-}}\left(\hat L^{+}_{g}
  -\hat L^{-}_{g}\right) d^4 x  \, .
\end{equation}
Here $\hat L^{+}_{g}$ and $\hat L^{-}_{g}$ are Lagrangians defined by the
relation (\ref{4.19}) and calculated as a limiting magnitude when
approaching the hypersurface $\Sigma^{(3)}_t$ from outside or inside
respectively. Under the infinitesimal normal displacement of the
hypersurface $\Sigma^{(3)}_t$, the full action is variated by the formula
\begin{eqnarray}\label{4.21}
    \delta I^{(g)}_{tot} =
   - \int\limits_{\Sigma^{(3)}_t}\left(\hat L^{+}_{g}
   -\hat L^{-}_{g}\right) \delta x^{\mu} d \Sigma_{\mu} =
   \int\limits_{\Sigma^{(3)}_t} [\hat L_{g}]\delta\lambda d\Sigma \, .
\end{eqnarray}
Hence, from arbitrariness of $\delta\lambda (p)$ and the requirement $
\delta I^{(g)}_{tot}=0$, one finds
\begin{equation}\label{4.22}
      [\hat L_{g}]=\hat L^{+}_{g} -\hat L^{-}_{g}
      =[K^{\mu}_{\nu}K^{\nu}_{\mu} - K^2 ]
      =2\bar K^{\mu}_{\nu}([K^{\nu}_{\mu}]-[K]\delta^{\nu}_{\mu})=0\, .
\end{equation}
Here we considered that $[^{(3)}R]=0$ on $\Sigma^{(3)}_t$. Then, using
(\ref{4.16}), from (\ref{4.22}) we obtain
\begin{equation}\label{4.23}
     \bar K_{\mu\nu} u^{\mu} u^{\nu} = 0.
\end{equation}

The relations (\ref{4.16}) and (\ref{4.23}) form the necessary complete
set of algebraic conditions or constraints for the extrinsic curvature
tensor $K^{\mu}_{\nu {|_{\pm}}}$ of the hypersurface $\Sigma^{(3)}_t$.

Now we can eliminate gravitational degrees of freedom in the action
$I^{(g)}_{tot}$ and construct the action for the shell. For this purpose
it is necessary to calculate $I^{(g)}_{tot}$ on the solutions of the
vacuum Einstein equations (\ref{4.13}) taking into account the constraints
(\ref{4.16}) and (\ref{4.23}). Note, first, that on this stage we use
explicitly only the following results of these equations:
\begin{eqnarray}\label{4.24}
      ~^{(4)}R &=& 0\, , \qquad
      [\omega^{\mu}]n_{\mu} =2[K] = \chi \sigma\, .
\end{eqnarray}
Substituting these relations for the corresponding terms in (\ref{4.1})
one finds
\begin{equation}\label{4.26}
     {I^{~(g)}_{tot}}_{| \{equations~(\ref{4.24})\}}
      = I_{sh} + I_{\partial D^{(4)}} + I_{0}\, ,
\end{equation}
where
\begin{equation}\label{4.27}
        I_{sh} =
      \frac{1}{2}\int\limits_{\Sigma^{(3)}_{t}}c\sigma d^3\Omega
        =-\frac{c}{2}\int\limits_{S^{(2)}_{t}} \ d^2 m~
    \int\limits_{\gamma}\omega
\end{equation}
is the reduced action for the dust shell. This action must be considered
together with constraints (\ref{4.16}) and (\ref{4.23}). The action
$I^{(g)}_{sh}$ is quite certain if the gravitational fields in the
neighbourhood of $\Sigma^{(3)}_{t}$ are determined as the solutions of the
vacuum Einstein equations (\ref{4.13}) which satisfy the boundary
conditions (\ref{4.16}) and (\ref{4.23}). That is the finding of these
fields that completes the construction of the action for the shell. At
this stage all the equations (\ref{4.13}) and constraints (\ref{4.16}),
(\ref{4.23}) are already used.

Note, that one usually comes to the action for the shell in the other
form. In our approach the action can be obtained at the partial reduction
of initial action $I^{(g)}_{tot}$, when the constraint in (\ref{4.24}) is
not taken into account. As a result we come to the action of the type
\begin{equation}\label{4.27a}
       \tilde{I}_{sh}  =
     - c\int\limits_{\Sigma^{(3)}_t}
     \left(\sigma - \frac{1}{\chi}[K]\right)\omega \wedge d^2\Omega\, .
\end{equation}
or to some its modification. In the spherically-symmetric case from here
follows the Lagrangian of the shell in a frame of reference of the
comoving observer. However quantity $[K]$ contains second derivatives with
respect to proper time of the shell. When eliminating them, through the
integration by parts, one comes to rather complicated Lagrangians and
Hamiltonians.

To find the equations of motion for particles of the shell from action
$I_{sh}$ (\ref{4.27}) one should introduce the independent coordinates
$x^{\mu}_{\pm}$ in each of the regions $D^{(4)}_{\pm}$, and the interior
coordinates $y^i \,(i,k=0,2,3)$ on $\Sigma^{(3)}_{t}$. Let the equations
of embedding of $\Sigma^{(3)}_{t}$ into $ D^{(4)}_{\pm}$ have the form
$x^{\mu}_{\pm}=x^{\mu}_{\pm}(y^i)$. Then we can write the relations
\begin{eqnarray}
    & ~^{(4)}ds^{2}_{\pm}
       =  g^{\pm}_{\mu\nu} dx^{\mu}_{\pm} dx^{\nu}_{\pm}\, , \qquad
      ~^{(3)}ds^{2} =
      g^{\pm}_{\mu\nu} x^{\mu}_{\pm,i} x^{\mu}_{\pm,k}dy^i dy^k
      = h_{ik}dy^i dy^k \, ,     \label{4.28} \\
   & \omega = \omega^{\pm} = u^{\pm}_{\mu}dx^{\mu}_{\pm}\, , \qquad
      \omega^{\pm}_{|\gamma} = ds_{\pm} \, , \qquad
      u_{\pm}^{\mu} = dx^{\mu}_{\pm}/ds_{\pm} \, , \label{4.29} \\
   & ~^{(3)} \omega
       = u^{\pm}_{\mu}x^{\mu}_{\pm,i}dy^i = u_{i}dy^i \, , \quad
       ~^{(3)} \omega_{|\gamma} = ~^{(3)}ds\, , \quad
       u^i = dy^{i}/~^{(3)}ds\, .         \label{4.30}
\end{eqnarray}
Non-gravitational interaction between particles of the dust is absent.
Therefore we consider quantity $d^2 m$ to be unchanged when a flow line is
varied.

First, consider variations $I_{sh}$ with respect to the internal
coordinates $y^i $. In this case $\int_{\gamma}\omega
=\int_{\gamma}\,^{(3)} \omega_{\,|\gamma}=\int_{\gamma}\,^{(3)}ds$. Then
the metric $h_{ik}(y^i)$ is given on $\Sigma^{(3)}_{t}$ and the variation
of $I_{sh}$ leads to the equations of three-dimensional geodesic on the
hypersurface $\Sigma^{(3)}_{t}$
\begin{equation}\label{4.31}
          u^{i}_{;\, k}u^{k} =0\, .
\end{equation}
Here``$~_{;k} $'' denotes the covariant derivative with respect to the
coordinate $y^k$ calculated with the help of the metric $h_{ik}$.

The consideration of the variational principle $\delta I^{(g)}_{sh}= 0$
with respect to the exterior coordinates $x^{\mu}_{\pm}$ is more
interesting treatment. In this case $\int_{\gamma}\omega =\int_{\gamma}\,
\omega^{\pm}_{|\gamma}=\int_{\gamma}^{(4)}ds_{\pm}\,$ Then the metrics
$g^{\pm}_{\mu\nu}(x^{\rho})$ are given in a neighbourhood of the shell.
Since the normal variations of the shell are already used, it is possible
to consider the variations of dynamical quantities, generated only by the
tangential to $\Sigma^{(3)}_{t}$ variations of the coordinates $x^{\mu}$.
These variations of the values will be denoted by the sign $\tilde \delta
$. Thus, omitting for simplicity signs ``$ \pm $'', we have
\begin{equation}\label{4.32}
       \tilde \delta x^{\mu} =
      \delta x^{\mu} + n^{\mu}n_{\nu}\delta x^{\nu}
      \equiv h^{\mu}_{\nu} \delta x^{\nu} \quad
      (n_{\mu} \tilde \delta x^{\mu} =0, \quad
      h^{\mu}_{\nu}=\delta^{\mu}_{\nu} +  n^{\mu}n_{\nu})\, ,
\end{equation}
where $ \delta x^{\mu} $ are arbitrary values. Then we find
\begin{equation}\label{4.33}
      \tilde \delta\omega_{|\gamma} =  \tilde \delta\, ^{(4)}ds =
      \tilde \delta \sqrt{g_{\mu\nu}dx^{\mu}dx^{\nu}}
      =- u_{\mu;\nu}u^{\nu}h^{\mu}_{\rho}\delta x^{\rho}\, ^{(4)}ds
      + d(u_{\mu} \delta x^{\mu})\, .
\end{equation}
Supposing that $\delta x^{\mu}=0 $ on $\Sigma^{(3)}_{t} \cap{\partial
D}^{(4)}$, from the requirement $\tilde\delta I^{(g)}_{sh}=0 $ we obtain
the three-dimensional geodesic equations on $\Sigma^{(3)}_{t} $, but,
here, in the four-dimensional form
\begin{equation}\label{4.34}
      u_{\mu;\nu}u^{\nu}h^{\mu}_{\rho} =0\, .
\end{equation}
This equation cab be rewritten as
\begin{equation}\label{4.34a}
   u_{\rho;\nu}u^{\nu} = - u_{\mu;\nu}u^{\nu}n^{\mu}n_{\rho} \, .
\end{equation}
Hence, using the definition of $K_{\mu\nu}$ (\ref{4.16a}) one obtains
\begin{equation}\label{4.35}
      u_{\rho;\nu}u^{\nu} = n_{\rho} n_{\mu;\nu}u^{\mu}u^{\nu}
            = - n_{\rho} K_{\mu\nu}u^{\mu}u^{\nu}\, .
\end{equation}

Here we again introduce signs ``$\pm $'' and use the relations $K_{\mu\nu
|_{\pm}}=\bar K_{\mu\nu}\pm\frac{1}{2}[K_{\mu\nu}]$. Then, taking into
account constraints (\ref{4.16b}) and (\ref{4.23}), we come to the
equations of motion for the shell's particles with respect to the exterior
coordinates
\begin{equation}\label{4.37}
      \left({u_{\mu;\nu}u^{\nu}}\right)_{|\pm} =
      \pm \frac{\chi}{4}\sigma n_{\mu}\, .
\end{equation}
For completeness one should add the unused constraints
\begin{equation}\label{4.38}
      [K_{\mu\nu}]u^{\mu}e^{\nu}_{i} = 0 \, , \qquad
      [K_{\mu\nu}]e^{\mu}_{a} e^{\nu}_{b} =\frac{\chi \sigma}{2}h_{ab}\, ,
\end{equation}
where $ e^{\mu}_{a} = \partial x^{\mu}/ \partial y^a $.

From (\ref{4.37}) it follows the well-known Israel equations
\cite{israel1}
\begin{eqnarray}
      n^{\mu}{\frac{Du_{\mu}}{ds}}\biggl |_{+} & + &
      n^{\mu}{\frac{Du_{\mu}}{ds}}\biggl |_{-}  =  0\, , \qquad
      e^{\nu}_{i}{\frac{Du_{\mu}}{ds}}\biggl |_{\pm} = 0\, , \label{4.39} \\
      n^{\mu}{\frac{Du_{\mu}}{ds}}\biggl |_{+} & - &
      n^{\mu}{\frac{Du_{\mu}}{ds}}\biggl |_{-} =
     - \frac{\chi \sigma}{2}\, ,      \label{4.40}
\end{eqnarray}
where $Du_{\mu}= u_{\mu;\nu}dx^{\nu}$ is the covariant differential.

The equations of motion of the dust shell (\ref{4.37}) can immediately be
found from the action $I_{sh}$. Indeed, acting in the same manner as when
deducing the equations of motion (\ref{4.37}), the variational formula
(\ref{4.33}) can be transformed to the form
\begin{equation}\label{4.41}
      \tilde \delta\, ^{(4)}ds_{|_{\pm}} =
      - u_{\mu;\nu}u^{\nu} \delta x^{\mu}_{|_{\pm}}\, ^{(4)}ds_{|_{\pm}}
    \mp \frac{1}{2}[K_{\mu\nu}]u^{\mu}u^{\nu}n_{\rho}\delta
    x^{\rho}ds_{|_{\pm}}
      + {d(u_{\mu} \delta x^{\mu})}_{|_{\pm}}\, .
\end{equation}
or
\begin{equation}\label{4.42}
    \tilde \delta\omega^{\pm}_{|\gamma} =
      \tilde \delta\, ^{(4)}ds_{|{_\pm}} = \left\{\left(- u_{\mu;\nu}u^{\nu}
      \pm \frac{\chi \sigma}{4} n_{\mu}\right)\delta x^{\mu}\, ^{(4)}ds
      + {d(u_{\mu} \delta x^{\mu})}\right\}_{|\pm}\, .
\end{equation}
From here, under the above conditions, the equations of motion follow.

The proposed variational deducing of the equations of motion makes the
problem of construction of the effective action for the dust shell free
from constrains (\ref{4.16}) and (\ref{4.23}). It turns out that it is
possible for some special class of the configurations. To show it, we
shall choose such interior coordinates $y^i $, which at $i=a = 2,3 $ are
the Lagrange coordinates of particles on the shell $\Sigma ^{(3)}_{t}$. In
addition, we introduce the coordinates $x^{\mu}_{|\pm}$ in the regions
$D^{(4)}_{\pm}$ so that, when $\mu=a=2,3$ the equalities
${x^{a}_{+}}|_{\Sigma^{(3)}_{t}}={x^{a}_{-}}|_{\Sigma^{(3)} _{t}}=y^a $
are satisfied. These coordinates are arbitrary in any other respect. Then
the formulas of embedding of $\Sigma^{(3)}_{t}$ into $D^{(4)}_{\pm}$ have
the form $x^{n}_{\pm}=x^{n}_{\pm}(y^0)\ (n=0,1)$ or $f_{\pm}(x^0, x^1)
=0$. Therefore we have $u^{\mu}_{\pm} =
\{u^{0}_{\pm},~u^{1}_{\pm},~0,~0\}$ and
$n_{\mu}^{\pm}=\{n_{0}^{\pm},~n_{1}^{\pm},~0,~0\}$. Using the conditions
$(u_{\mu}u^{\mu})_{|\pm} =- (n_{\mu}n^{\mu})_{|\pm} =1 $ and
$({u_{\mu}n^{\mu}})_{|\pm} = 0$ one finds $ n_{0}^{\pm}= u^{1}_{\pm}\,
,\quad n_{1}^{\pm}=- u^{0}_{\pm}\, $. Hence it follows
\begin{equation}\label{4.44}
      n_{\mu}\delta x^{\mu}ds_{_{|\pm}} =
      {(u^1 \delta x^0 -u^0 \delta x^1)ds}_{_{|\pm}}
      = {(dx^1 \delta x^0 -dx^0 \delta x^1)}_{_{|\pm}}\, .
\end{equation}
Therefore the variational formula (\ref{4.42}) has the form
\begin{equation}\label{4.45}
       \tilde \delta\omega^{\pm}_{|\gamma} =
       \tilde \delta\, ^{(4)}ds_{|_{\pm}} = \left\{\delta ^{(4)}ds
     \pm \frac{1}{4} \chi \sigma {(dx^1 \delta x^0 - dx^0 \delta x^1)}
      + d(u_{\mu} \delta x^{\mu})\right \}_{|\pm}\, .
\end{equation}

Now we introduce the vector potential $U_{n}=U_{n}(x^0, x^1)$ by the
relation
\begin{equation}\label{4.46}
      d\wedge (U_{n}dx^n ) \equiv G_{01} dx^0 \wedge dx^1 =
      -\frac{1}{4}\chi\sigma dx^0 \wedge dx^1 \, ,
\end{equation}
where $G_{nm}\equiv U_{m,n}-U_{n,m} \ (n,m=0,1)$. Hence it follows that
the configurations, being considered, admit such motions of matter for
which $\sigma = \sigma (x^0, x^1)$.

Using the definition (\ref{4.46}) and the relation
\begin{equation}\label{4.48}
  \delta (U_{n}dx^n ) -
  d(U_{n} \delta x^n)= G_{10}(dx^0 \delta x^1 -dx^1 \delta x^0) \, ,
\end{equation}
the variational formula (\ref{4.45}) can be rewritten in the following
form
\begin{eqnarray}\label{4.49}
      \tilde \delta\omega^{\pm}_{|\gamma} =
       \tilde \delta\, ^{(4)}ds_{|_{\pm}}
       =  \left\{\delta (ds \mp U_{n}dx^n )
      + d[(u_n \pm U_n)\delta x^{n} + u_a \delta y^a] \right \}_{|\pm}\, .
\end{eqnarray}
Returning to action for the shell (\ref{4.27}), we conclude, that in the
case under consideration we have
\begin{equation}\label{4.50}
       \delta I_{sh}
       = \delta I^{\pm}_{sh} - \frac{c}{2}\int\limits_{S^{(2)}_{t}}d^2 m
       \biggl \{ (u_n \pm U_n) x^{n}_{,\,0}\delta y^0
      + u_a \delta y^a \biggl \}_{\pm}\ \biggl |^{B}_{A} \, ,
\end{equation}
where
\begin{equation}\label{4.51}
      I^{\pm}_{sh} = -\frac{c}{2}\int\limits_{S^{(2)}_{t}} d^2 m
   \int\limits_{\gamma} \left(ds \mp U_{n}dx^n \right)_{|\pm}\, ,
    \qquad (n=0,1)
\end{equation}
is the effective action for the shell written in terms of the exterior
coordinates. Indices $A$ and $B$ indicate that the corresponding
quantities are taken in initial and final positions of the shell. Since at
fixed initial and final positions of particles ${\delta y^i}|_{A,B}=0$,
then it follows $\delta I_{sh}=\delta I^{\pm}_{sh}$.

In such away, under the above conditions, the action of the shell
(\ref{4.27}) with the constraints (\ref{4.16}), (\ref{4.23}) and the
action (\ref{4.50}) without these constraints are equivalent. The actions
$ I^{+}_{sh}$ and $ I^{-}_{sh}$ are equivalent in the same sense. Let us
note, that in the considered above independent treatment of the interior
and exterior faces of the shell there are new constraints following from
isometry conditions of these faces.

\section{Effective action for the spherical dust shell}

Let us consider spherically-symmetric compound region $D^{(4)}=D^{(4)}_{-}
\cup \Sigma^{(3)}_{t} \cup D^{(4)}_{+}\subset V^{(4)}$ into the
spherically-symmetric space-time $V^{(4)}$, where $D^{(4)}_{\mp}$ are
exterior and interior regions separated from each other by
spherically-symmetric time-like hypersurface $\Sigma^{(3)}_{t}$. By using
the curvature coordinates we can choose common in $D^{(4)}_{\pm}$,
spatial, spherical coordinates $\{r,~\theta,~\alpha\}$, and individual
time coordinates $t_{\pm}$ for $D^{(4)}_{\pm}$ respectively. Then the
world sheet for the shell $\Sigma^{(3)}_{t}$ respectively the interior and
exterior coordinates is determined by the equations $r=R_{-}(t_{-})$ and
$r=R_{+}(t_{+})$. Under appropriate choice of $t_{\pm}$ we have
$R_{-}(t_{-})=R_{+}(t_{+})$. Thus, the interior and exterior regions are
determined by the relations $$ D^{(4)}_{-}=\{t_{-},~r,~\theta,~\alpha
:~r_{0} <r<R_{-}(t_{-})\}, \quad D^{(4)}_{+}=\{t_{+},~r,~\theta,~\alpha
:~R_{+}(t_{+})<r<\infty\}$$ for all $\{\theta, \alpha \} \in \{0\leq
\theta\leq \pi,\ 0\leq\alpha <2\pi ,\}$ and for all admissible $t_{\pm}$.
The particles of the shell are described by one collective dynamical
coordinate $R=R_{\pm}(t_{\pm})$ and by the two fixed individual (Lagrange)
angular coordinates $\theta$ and $\alpha$. The minimal value of $r_{0} $
is limited by the domain of definition of the curvature coordinates.

The gravitational fields into the regions $D^{(4)}_{\pm}$ are given by the
metrics
\begin{equation}\label{4.2.1}
      ~^{(4)}ds^{2}_{\pm}=f_{\pm}c^2 dt^{2}_{\pm}- f^{-1}_{\pm}dr^2
      -r^2 (d\theta^2+\sin^2 \theta d\alpha^2)\, ,
\end{equation}
where
\begin{equation}\label{4.2.1a}
      f_{\pm} = 1 - \frac{2\gamma M_{\pm}}{c^2 r}\, ,
\end{equation}
and $M_{\pm}$ are the Schwarzschild masses $(M_{+}>M_{-})$.

Owing to the spherical symmetry $\sigma =\sigma(t_{\pm},R)$. Therefore the
conditions of applicability of the modified action (\ref{4.50}) are
satisfied. In this case we have
\begin{eqnarray}
     & d^2 m = \sigma d^2 \Omega =
      \sigma R^2 \sin \theta d\theta d\alpha\, , \label{4.2.2}  \\
     & U_{n}dx^n=c\varphi(t_{\pm},R)dt_{\pm}
      + U_{R}(t_{\pm},R)dR \, .    \label{4.2.3}
\end{eqnarray}
Using the gauge condition $U_{R}(t_{\pm},r)=0 $, the action (\ref{4.50})
can be written in the form
\begin{equation}\label{4.2.4}
      I^{\pm}_{sh} =
      -\frac{c}{2}\int\limits_{S^{(2)}_{t}}
      \sigma R^2 \sin \theta d\theta d\alpha
      \int\limits_{\gamma_{\pm}} \left(^{(2)}d s \mp c \varphi d t
      \right)_{|\pm}\, .
\end{equation}
Since the particles move only radially $(\theta =\mbox{const}, \ \varphi
=\mbox{const})$ we shall use the truncate interval
\begin{equation}\label{4.2.5}
      ~^{(2)}ds^{2}_{\pm}=f_{\pm}c^2 dt^{2}_{\pm}- f^{-1}_{\pm}dR^2\, .
\end{equation}
Further, from the formula (\ref {4.46}) it follows
\begin{equation}\label{4.2.6}
       \frac{1}{4} \chi \sigma = \frac{\gamma m}{2c^2 R^2}
      =\frac{\partial \varphi}{\partial R}\, ,
\end{equation}
where $m = 4\pi\sigma R^2$ is the rest mass of the shell. Hence, up to an
additive constant, one finds
\begin{equation}\label{4.2.7}
      \varphi = - \frac{\gamma m}{2c^2 R}\, .
\end{equation}
Finally, integrating in (\ref{4.2.4}) over the angles $\theta$ and
$\alpha$ and making use of (\ref{4.2.5}) and (\ref{4.2.7}), the effective
action of the shell can be expressed in the form
\begin{equation}\label{4.2.8}
   I^{\pm}_{sh} = \frac{1}{2}\int\limits_{\gamma_{\pm}}L^{\pm}_{sh}dt_{|\pm}
      =-\frac{1}{2}\int\limits_{\gamma_{\pm}}
      \biggl(mc~^{(2)}ds \pm \frac{\gamma m^2}{2 R} dt\biggr)_{|\pm}\, ,
\end{equation}
where
\begin{equation}\label{4.2.9}
      L^{\pm}_{sh} =
      -mc^2 \sqrt{f_{\pm}- f^{-1}_{\pm} R^2_{t\pm}/c^2}
      \pm U
\end{equation}
are the Lagrangians of the dust shell with respect stationary observes
into the regions $D^{(4)}_{\pm}\, , \ (R_{t\pm}=dR/dt_{\pm})$, and
\begin{equation}\label{4.2.10}
      U^{(G)} = - \frac{\gamma m^2}{2R}
\end{equation}
is the effective potential energy of the gravitational self-action of the
shell. It is important that the self-action (\ref{4.2.10}) has the same
form as that in the Newtonian theory (formula (\ref{2.4}) in Appendix C).
The Lagrangians (\ref{4.2.9}) themselves can be obtained from the
corresponding Newtonian analogs (see Appendix C, formulas (\ref{2.5}) and
(\ref{2.6})) by the formal replacement of the first and second terms
describing the kinetic and potential energies of the shell into the
external field by their general relativistic analog, the geodesic
Lagrangian $-m c~^{(2)}ds_{\pm}/dt_{\pm}$. It is natural that the
Lagrangians (\ref{2.5}), (\ref{2.6}), up to an additive constant, are the
Newtonian limits of the relativistic Lagrangians (\ref{4.2.9}).

It is easy to see that the actions (\ref{4.2.8}) transform each into other
under the discrete gauge transformation $$
      M_{\pm} \stackrel{~}{\to} M_{\mp} \quad
      (f_{\pm}\stackrel{~}{\to}f_{\mp}), \quad
      U^{(G)}\stackrel{~}{\to} - \ U^{(G)}, \quad
      t_{\pm}\stackrel{~}{\to} t_{\mp}.
$$ This transformation generalize the corresponding transformation of the
Newtonian theory of shells (see Appendix C) and reduce to the
transformation from the interior observer to the exterior one and
otherwise.

Note, that despite the equivalence of the actions $I^{\pm}_{sh} $, similar
to Newtonian case (\ref{2.5}), (\ref{2.6}), they can be considered quite
independently. We also can consider the regions $D^{(4)}_{\pm}$ together
with the corresponding gravitational fields (\ref{4.2.1}) separately and
independently, as manifolds with the edge $\Sigma^{(3)}_{t\pm}$. The edges
$\Sigma^{(3)}_{t\pm}$ acquire the physical meaning of different faces of
the shell with the world sheet $\Sigma^{(3)}_{t}$, provided the regions
$D^{(4)}_{\pm}$ are joined along these edges $\Sigma^{(3)}_{t\pm}$. This
can be performed only if the conditions of isometry of the edges
$\Sigma^{(3)}_{t\pm}$ are satisfied
\begin{equation}\label{4.2.11}
      f_{+}c^2 dt^{2}_{+}- f^{-1}_{+}dR^2 =
      f_{-}c^2 dt^{2}_{-}- f^{-1}_{-}dR^2 = c^2 d \tau^2 \, ,
\end{equation}
where $\tau$ is the proper time of the shell. In addition we have
$\Sigma^{(3)}_{t+}=\Sigma^{(3)}_{t-}=\Sigma^{(3)}_{t}$, $\gamma_{+}(t_{+})
=\gamma_{-}(t_{+})=\gamma$.

Now we study some results following from the isometry conditions of the
edges. First, we obtain the relations between the velocities
\begin{equation}\label{4.2.12}
     c^2 \frac{f_+}{R^2_{t+}} - \frac{1}{f_+} =
      c^2 \frac{f_-}{R^2_{t-}} - \frac{1}{f_-}\, ,
\end{equation}
\begin{equation}\label{4.2.13}
     R_{\tau}^2 \equiv \left(\frac{dR}{d\tau}\right)^2 =
     \frac{c^2 R^2_{t\pm}}{c^2 f_{\pm} - f^{-1}_{\pm} R^2_{t\pm}}\, , \qquad
     R_{t\pm}^2 \equiv \left(\frac{dR}{dt_{\pm}}\right)^2
     = \frac{c^2 f^{2}_{\pm} R^{2}_{\tau}}{c^2 f_{\pm} + R^2_{\tau}} \, .
\end{equation}
Then from the Lagrangians $L^{\pm}_{sh} $ (\ref{4.2.9}) one finds the
momenta and Hamiltonians of the shell
\begin{eqnarray}
      P_{\pm} = \frac{\partial L^{\pm}_{sh}}{\partial R_{t\pm}}
      = \frac{mR_{t\pm}}
      {f_{\pm} \sqrt{f_{\pm} - f^{-1}_{\pm} R^2_{t\pm}/c^2}}
      = \frac{m}{f_{\pm}} R_{\tau}\, , \label{4.2.14}
\end{eqnarray}
\begin{eqnarray}
      H^{\pm}_{sh} = \frac{mc^2 f_{\pm}}
      {\sqrt{f_{\pm} - f^{-1}_{\pm} R^2_{t\pm}/c^2}}
      \mp U = mc^2 f_{\pm} \frac{dt_{\pm}}{d\tau}
      \mp U     \label{4.2.15}
\end{eqnarray}
or
\begin{eqnarray}\label{4.2.16}
      H^{\pm}_{sh} = c \sqrt{f_{\pm}(m^2 c^2 + f_{\pm}P_{\pm}^2)}
      \mp U  =
      mc^2 \sqrt{f_{\pm} + R^2_{\tau}/c^2} \mp U = E_{\pm}\, ,
\end{eqnarray}
where $E_{\pm}$ are the energies of the shell which are conjugated to the
time $t_{\pm}$ respectively and conserve with respect to the corresponding
interior or exterior stationary observers' point of view. After
elimination of velocity $R_{\tau}$ from (\ref{4.2.14}) and (\ref{4.2.16}),
the isometry conditions of the edges can be expressed in the form
\begin{eqnarray}
      & f_{+} P_{+} = f_{-} P_{-} \, , \label{4.2.14a}\\
      & \left(E_{-}-U\right)^2 -m^2 c^4 f_{-}
        = \left(E_{+}+U\right)^2 -m^2 c^4 f_{+}\, .\label{4.2.17}
\end{eqnarray}
Substituting $U$ and $f_{\pm}$ from (\ref{4.2.1a}) and (\ref{4.2.10}) for
those in the last relation and equating the coefficients at the same power
of $R$ we obtain the relations between the Hamiltonian $H^{\pm}_{sh}$ and
the Schwarzschild masses $M_{\pm}$
\begin{equation}\label{4.2.18}
      H^{+}_{sh}= H^{-}_{sh} = (M_{+} - M_{-})c^2 = E \, .
\end{equation}
Here $E=E_{\pm}$ is the full energy of the shell. This energy is
conjugated to the coordinate time $t_{+}$ and $t_{-}$ as well, and does
not depend on the position of the stationary observer (inside or outside
the shell). We shall interpret the relations (\ref{4.2.14a}) and
(\ref{4.2.18}) following from the above independent consideration of the
shell faces, as momentum and Hamiltonian constraints.

The Lagrangians $L^{\pm}_{sh}$ (\ref{4.2.9}), as well as the relations
(\ref{4.2.12}) - (\ref{4.2.18}), are valid only in a limited domain, since
the used curvature coordinates are valid outside the event horizon only.
Therefore $L^{-}_{sh}$ can be used when $R>2\gamma M_{-}/c^2 $, and
$L^{+}_{sh}$ when $R>2\gamma M_{+}/c^2 \ \ (M_{+}>M_{-})$.

As is known, the complete description of the shells can be performed in
the Kruskal-Szekeres coordinates. With respect to these coordinates the
full Schwarzschild geometry consists of the four regions $R^{+},\ T^{-},\
R^{-}, T^{+}$, detached by the event horizons. Our above consideration
concerned with the  $R^{+}$ region only.

Supposing $r$ to be a time coordinate, we can formally use the action in
the form (\ref{4.2.8}) under the horizon. However, here we encounter the
ambiguity when choosing the sign before $~^{(2)} d s $. It is usually
ascribed to ambiguity of the radial component direction of the unit normal
to $\Sigma^{(3)}_{t}$. The point is that in the curvature coordinates the
regions $T^{-}$ and $T^{+}$ coincide. Hence the time singularity $r=0 $
contains the two singularities: past singularity and future singularity.
Therefore, for instance, the movement of a test particle with the energy
$E=0$ consists of the two stages. At the first stage the particle begins
to move into the expanding region $T^{+}$ from the past singularity $r=0$
and reaches the horizon at a moment when $r$  reaches $2\gamma M/c^2 $.
Then it goes over into the contracting $T^{-}$ region and moves from the
horizon to the future singularity $r=0$. In the coordinates $\{r,t\}$,
where $r$ is the time coordinate, the latter stage looks like the movement
directed backwards in time.

Similarly, in the curvature coordinates the regions $R^{-}$ and $R^{+}$ of
the Kruskal-Szekeres diagram coincide and ordinary movement of particles
into the future of the $R^{-}$-region looks as the movement directed
backwards in time which corresponds to the change $d s \stackrel{~}{\to}-d
s $. It means that ordinary particles moving into the $R^{-}$-region are
mapped into the $R^{+}$-region as antiparticles (remember the Feynman's
interpretation of antiparticles as ordinary particles moving backwards in
time). Such trajectories can be taken into account by the change of the
sign before $m c~^{(2)}d s_{\pm}$ in the expression for the action
(\ref{4.2.8}) of the shell.

In order to use simplicity and convenience of the curvature coordinates
and, at the same time, to keep information about shells into the
$R^{-}$-region we introduce an auxiliary discrete variable $\varepsilon
=\pm 1 $ and make a change $~^{(2)}ds_{\pm} \stackrel{~}{\to}
\varepsilon_{\pm}~^{(2)} d s_{\pm} $ in $I^{\pm}_{sh} $ (\ref{4.2.8}).
Herewith $\varepsilon_{\pm} =1 $ correspond to the shell into the
$R^{+}$-region, and $\varepsilon_{\pm} =-1 $ to the shell into the
$R^{-}$-region. Then, we introduce the quantities
$\mu_{\pm}=\varepsilon_{\pm}m$. As a result the extended action has the
form
\begin{eqnarray}\label{4.2.19}
   I^{\pm}_{sh}(\mu_{\pm}) =
      \frac{1}{2}\int\limits_{\gamma_{\pm}}L^{\pm}_{sh}(\mu_{\pm}) dt_{|\pm}
      =-\frac{1}{2}\int\limits_{\gamma_{\pm}}
      \left(\mu c~^{(2)}ds \mp U^{G} dt\right)_{|\pm}\, ,
\end{eqnarray}
where
\begin{equation}\label{4.2.20}
      L^{\pm}_{sh}(\mu_{\pm}) =
      - \mu_{\pm} c^2 \sqrt{f_{\pm}- f^{-1}_{\pm} R^2_{t\pm}/c^2}
      \pm U
\end{equation}
are the generalized Lagrangians describing the shell inside any of the
$R^{\pm}$-regions with respect to the curvature coordinates of the
interior $\{ t_{-}, R \}$ or exterior $\{ t_{+}, R \}$ regions. The event
horizons $R_g=2\gamma M_{\pm}/c^2$ are, still, singular points of the
dynamical systems (\ref{4.2.19}) and must be excluded from consideration.

For the extended system (\ref{4.2.19}) the Hamiltonian has the form
\begin{equation}\label{4.2.21}
      H^{\pm}_{sh}(\mu_{\pm}) =
      c\varepsilon_{\pm} \sqrt{f_{\pm}(m^2 c^2 + f_{\pm}P_{\pm}^2)}
      \mp U =
      \mu_{\pm}c^2 \sqrt{f_{\pm} + R^2_{\tau}/c^2} \mp U \, .
\end{equation}
Hence, taking into account the Hamiltonian constraints (\ref{4.2.18}) one
finds the standard relations of the theory of dust spherical shells
\cite{israel1}. We shall rewrite them in terms of new designations
\begin{eqnarray}
      \mu_{-} \sqrt{f_{-} + R^2_{\tau}/c^2} &-&
      \mu_{+} \sqrt{f_{+} + R^2_{\tau}/c^2} =
      \frac{\gamma \mu^2}{Rc^2}\, ,   \label{4.2.22a}  \\
      \mu_{-} \sqrt{f_{-} + R^2_{\tau}/c^2} &+&
      \mu_{+} \sqrt{f_{+} + R^2_{\tau}/c^2} =
      2(M_{+}-M_{-})\, .  \label{4.2.22b}
\end{eqnarray}
In the end of the section we write out the Hamilton-Jacobi equations
corresponding to the Hamiltonians (\ref{4.2.21}) and to the constraints
(\ref{4.2.14a}), (\ref{4.2.18}) for truncated actions
$S^{\pm}_{0}=S^{\pm}_{0}(R)$
\begin{equation}
      \frac{1}{f_{\pm}} \left(M_{+} -M_{-} \mp \frac{U}{c^2}\right)^2 -
      \frac{f_{\pm}}{c^2}\left(\frac{d S^{\pm}_{0}}{d R}\right)^2
      = m^2  \, . \label{4.2.22c}
\end{equation}
\begin{equation}\label{4.2.22d}
     f_{+}\, d\, S^{+}_{0} = f_{-}\, d\, S^{-}_{0} \, .
\end{equation}
Then, the complete actions are determined by the formula $S^{\pm}=-c^2
(M_{+}-M_{-})t_{\pm}+S^{\pm}_{0}$.

\section{Particular cases of spherical dust configurations}

$ \bullet $ {\bf Self-gravitating dust shell.} In this case $M_{-}=0$.
Denote $M_{+}= M$ and consider the shell moving into the $R_+ $-region.
Then with respect to the exterior coordinates, the Lagrangian and the
Hamiltonian of the shell have the form
\begin{equation}\label{4.2.23l}
      L^{+}_{sh} =
      -mc^2 \sqrt{1-\frac{2\gamma M}{c^2 R}-
      \left(1-\frac{2\gamma M}{c^2 R}\right)^{-1} \frac{R^2_{t+}}{c^2}}
      -  \frac{\gamma m^2}{2R} \, ,
\end{equation}
\begin{equation}\label{4.2.23f}
      H^{+}_{sh} = c~\sqrt{1-\frac{2\gamma M}{c^2 R}}\
      \sqrt{m^2 c^2 + \biggl(1-\frac{2\gamma M}{c^2 R}\biggr)P_{+}^2}
      + \frac{\gamma m^2}{2R} \, .
\end{equation}
The same shell with respect to the interior coordinates is described by
the Lagrangian and the Hamiltonian
\begin{equation}\label{4.2.23k}
      L^{-}_{sh} = - mc^2 \sqrt{1- R^2_{t-}/c^2 } + \frac{\gamma m^2}{2R} \, ,
\end{equation}
\begin{equation}\label{4.2.23g}
      H^{-}_{sh} = c\,\sqrt{m^2 c^2 + P_{-}^2}
                 - \frac{\gamma m^2}{2R} \, .
\end{equation}
This Hamiltonian was considered in the works  \cite{hajicek5},
\cite{dolgov}. The dynamical systems with $L^{\pm}_{sh}$ obey momentum and
Hamiltonian constraints $P_{-}= f_{+}P_{+}\,$,\ $ H^{+}_{sh} = H^{-}_{sh}
= Mc^2$ and they are canonically equivalent (see Appendix B).

$ \bullet$ {\bf The dust shell with vanishing full energy.} Now we
consider the shell for which the binding energy $E_b =(m + M_{-}-
M_{+})c^2 $ coincides with the rest energy $mc^2$. Denote $M_{+}=M_{-}
\equiv M$, $f_{+}=f_{-} \equiv f=1-2\gamma M/c^2 R$, $t_{+}=t_{-}\equiv t
$. Then for the full energy we have $E=0$. This is possible, as it follows
from (\ref{4.2.22a}), (\ref{4.2.22b} ), only when $\mu_{+}=-\mu_{-}<0 $,
i.e. for the wormhole. Such a shell can be considered as a classical model
for ``zeroth oscillations'' of dust matter with bare mass $m$ under the
gravitational field with $f=1-2\gamma M/c^2 R$.

In terms $\{t,~R\}$ the trajectories of ``zeroth oscillations'' are
determined by the equation
\begin{equation}\label{4.2.25}
      \frac{dR}{dt} = \frac{2c^3}{\gamma m}
      \left(1-\frac{2\gamma M}{c^2 R}\right)
      \sqrt{ \frac{\gamma^2 m^2}{4c^4}+\frac{2\gamma M}{c^2} R - R^2} \, .
\end{equation}
Hence for the turning radius we have
\begin{equation}\label{4.2.26}
      R_m = \frac{\gamma}{c^2}\left(
      M + \sqrt{M + \frac{m^2}{4}}~ \right)\, .
\end{equation}
In the case of the flat space when $M=0 $, from (\ref{4.2.25})and
(\ref{4.2.26}) we find
\begin{eqnarray}\label{4.2.27}
      \frac{dR}{dt} = c\,\sqrt{1-\frac{R^2}{R^2_{m0}}}\, ,\quad
      R_{m0}=\frac{\gamma m}{2c^2}\, .
\end{eqnarray}
The equations of motion of such ``zeroth'' shells coincide with those for
the oscillator
\begin{equation}\label{4.2.29}
         \frac{d^2 R}{dt^2} + \omega^2 R = 0 \, .
\end{equation}
Its oscillations $R(t)=R_{m0}\cos\omega(t-t_0)$ occur with the amplitude
$R_{m0}$ and frequency $ \omega = c/R_{m0} = 2 c^3/\gamma m$. Hence we
find the time of life of these shells into the flat space-time as a
half-period of the oscillation
\begin{equation}\label{4.2.31}
      T=\frac{\pi}{\omega} =
      \frac{\pi\gamma m}{2c^3} = \frac{\pi}{c}R_{m0}\, .
\end{equation}
For the shell with mass equal to the mass of the Earth we have $R_g
=2\gamma M/c^2\simeq 4cm$, $R_{m0}=R_g /4 \simeq 1cm, \ T \simeq 10^{-10}
c$. For the shells with Planck's mass $m=m_{pl}=\sqrt{\hbar c/\gamma}$ the
time of life equals $T=\pi T_{pl}/2$, where $T_{pl}=\sqrt{\hbar
\gamma/c^5}$ is Planck's time. We underline, that the ``zero'' shells are
characterized by that their gravitational binding energy completely
compensate proper energy, leaving their total energy to be equal to zero.
These shells can be thought of as a classical prototype of the Wheeler's
space-time foam \cite{wheeler}.

$ \bullet$ {\bf The set of concentric dust shells.} Now, consider briefly
configurations consisting from the set of $N$ concentric dust shells. Let
$R_a ,~m_a ,~\tau_a $ be the radius, bare mass and proper time of an
$a$-th shell, respectively $(a = 1, 2,...,N)$. For simplicity we suppose
that $ R_a > R_b $ if $ a > b $. Then let $M_{a}$ be the Schwarzschild
mass determining the gravitational field  $f_{a}=1-2\gamma M_{a}/c^2 r$ on
the right side of an $a$-th shell, into the region $R_a <r<R_{a+1}$.
Suppose $f^{-}_{a} = 1 - 2\gamma M_{a-1}/c^2 R_a$ and $f^{+}_{a} = 1 -
2\gamma M_{a}/c^2 R_a$. Let $ P^{\pm}_{a}=m_a dR_a / f^{\pm}_{a}d\tau_a $
be momenta of $a$-th shell, and $U^{(G)}_{a} = - \gamma m^{2}_{a}/2R_{a}$
be its potential energy of the self-action. Then
\begin{equation}\label{4.2.32}
      H^{\pm}_a =
      c\varepsilon^{\pm}_{a} \sqrt{f^{\pm}_{a}\left(m^2_{a} c^2
      + f^{\pm}_{a}(P^{\pm}_{a})^2\right)} \mp U_{a}
\end{equation}
is the Hamiltonians of an $a$-th shell. They, similarly to momenta
$P^{\pm}_{a}$, are considered from the stationary observers' points of
view, into the interior $R_{a-1} <r<R_{a}$ and exterior $R_a <r<R_{a+1}$,
regions respectively. They satisfy the momentum and Hamiltonian
constraints
\begin{equation}
        f^{+}_{a}P^{+}_{a}= f^{-}_{a}P^{-}_{a}\, , \qquad
      H^{+}_a =H^{-}_a = (M_a - M_{a-1} ) c^2 \, . \label{4.2.33}
\end{equation}
Now we are ready to determine the full Hamiltonian of this configuration
\begin{equation}\label{4.2.34}
      H_{N}= \sum^{N}_{a=1} H^{\pm}_{a}\, .
\end{equation}
For the self-gravitating configuration $M_{0}=0$. Then $H^{\pm}_1 = M_1
c^2 $ and the full Hamiltonian of the configuration satisfies the
constrain
\begin{equation}\label{4.2.35}
      H_{N}= Mc^2 \, .
\end{equation}
Here $M=M_N $ is the Schwarzschild mass of the configuration. The system
admits the discrete gauge transformations $$ M_{a} \leftrightarrow
M_{a-1}, \quad U_{a} \leftrightarrow -\, U_{a},\quad t_{a} \leftrightarrow
t_{a-1} \qquad (a = 1, 2,...,N)\, , $$ where $t_{a}$ is coordinate time
determined on the right from an $a$-th shell. The choice of sides (left or
right) of the shells is not fixed beforehand and can be made by the reason
of convenience.

\begin{acknowledgments}
I would like to acknowledge M.Korkina and S.Stepanov for helpful
discussions of problems, touched in this paper.
\end{acknowledgments}

\appendix
\section{Transformations of the surface equations}
We show that the surface equations (\ref{4.14}) reduce to the known
equations for the jumps of the extrinsic curvature tensor on the shell
\cite{israel1}. First, we shall calculate $ n_{\mu}[\omega^{\mu}]$.

We suppose that the following conditions are satisfied on the hypersurface
$\Sigma^{(3)}_{t} $
\begin{equation}\label{a1}
      [n_{\mu}] =0, \quad
      [n_{\mu,\nu}]h^{\nu}_{\sigma} =0, \quad
      [n^{\mu}_{,\nu}]h^{\nu}_{\sigma} =0, \quad
      [g_{\mu\nu,\rho}]h^{\rho}_{\sigma} =0\, .
\end{equation}
Hence it follows
\begin{equation}\label{a2}
      [\Gamma^{\mu}_{\rho\sigma}]
      h^{\nu}_{\mu}h^{\rho}_{\beta}h^{\sigma}_{\alpha}=0\, .
\end{equation}
Then from the definition (\ref{4.16a}) one finds
\begin{eqnarray}
      & [ \Gamma^{\sigma}_{\alpha\nu} ]n_{\sigma}h^{\nu}_{\beta}
       = [K_{\alpha\beta}] \, , \qquad
       [ \Gamma^{\sigma}_{\alpha\nu} ] n_{\sigma}h^{\alpha\nu}
       =[K] \, , \\ \label{a3}
      & [ \Gamma^{\sigma}_{\alpha \nu} ] n^{\alpha}h^{\nu}_{\beta}
       = - [K^{\sigma}_{\beta}] \, , \qquad
       [ \Gamma^{\sigma}_{\alpha\nu} ] n^{\alpha}h^{\nu}_{\sigma}
       =- [K] \, , \\ \label{a4}
      & [ \Gamma^{\sigma}_{\alpha\nu} ]
        n_{\sigma}n^{\alpha}h^{\nu}_{\beta} =0 \, . \label{a5}
\end{eqnarray}
According to (\ref{2.4}) we find  $$ n_{\mu}\omega^{\mu}=
      n_{\mu}g^{\sigma\rho}\Gamma^{\mu}_{\sigma\rho}-
      n^{\rho}\Gamma^{\sigma}_{\rho\sigma}=
      n_{\mu}h^{\sigma\rho}\Gamma^{\mu}_{\rho\sigma}-
      n^{\mu}h^{\sigma}_{\rho}\Gamma^{\rho}_{\mu\sigma}\, .  $$
Therefore, making use of Eq. (\ref{a3}) and (\ref{a5}) we obtain the
sought result (\ref{4.5a}).

Then, projecting the equation (\ref{4.14}) into the hypersurface
$\Sigma^{(3)}_{t} $ and into the normal $n^{\rho}$ one finds
\begin{eqnarray}
      Q_{\sigma\rho}n^{\rho}-\frac{1}{2}Qn_{\sigma} &=& 0\, ,\label{a7} \\
      Q_{\sigma\rho} h^{\sigma}_{\alpha} h^{\rho}_{\beta}
      -\frac{1}{2}Qh_{\alpha\beta}
      &=& - \chi\sigma u_{\alpha}u_{\beta}\, . \label{a8}
\end{eqnarray}
Using the definitions (\ref{4.15}) and Eqs. (\ref{a2})--(\ref{a5}) we
obtain
\begin{eqnarray}
    &  Q = g_{\mu\nu}  Q^{\mu\nu} =n_{\mu}[\omega^{\mu}] =2[K] \, ,\\
    &  Q_{\sigma\rho}n^{\rho} = n_{\mu}n^{\rho}[\Gamma^{\mu}_{\rho\sigma}]
       - \frac{1}{2}n_{\sigma}n^{\rho}[\Gamma^{\mu}_{\rho\mu}]
      +  \frac{1}{2}[\Gamma^{\mu}_{\sigma\mu}] =     \nonumber \\
    &  - n_{\sigma}(n_{\mu}[\Gamma^{\mu}_{\alpha\beta}]n^{\alpha}n^{\beta}
      + [\Gamma^{\mu}_{\mu\alpha}]n^{\alpha}) =
      - n_{\sigma}n^{\rho}[\Gamma^{\mu}_{\rho\nu}] h^{\nu}_{\mu}
      = [K] n_{\sigma}\, , \\
    &  Q_{\sigma\rho} h^{\sigma}_{\alpha} h^{\rho}_{\beta} =
      n_{\mu} [\Gamma^{\mu}_{\sigma\rho}]
      h^{\sigma}_{\alpha} h^{\rho}_{\beta} = [K_{\alpha\beta}]\, .
\end{eqnarray}
Thus the equation (\ref{a7}) is satisfied identically, and the equations
(\ref{a8}) yield the sought relations (\ref{4.16}).

\section{On the canonical equivalence of the actions $I^{\pm}_{sh}$
for the dust spherical shell}

In order to show the canonical equivalence of the actions $I^{\pm}_{sh}$
in the extended phase space we write the variational principle (\ref
{4.2.8}) in the form
\begin{equation}\label{b1}
       \delta I^{\pm}_{sh} = \delta\int (P_{\pm}dR - H_{\pm}dt_{\pm}) = 0\, ,
\end{equation}
where
\begin{equation}\label{b2}
   P_{\pm}=\frac{1}{cf_{\pm}}\sqrt{(H_{\pm}{\pm}U)^2 - m^2 c^4 f_{\pm}}\, .
\end{equation}
The dynamical systems with the actions $I^{\pm}_{sh}$ are restricted by
momentum and Hamiltonian constraints (\ref{4.2.14a}) and (\ref{4.2.18}),
which follow from the independent consideration of the faces of the shell.

The systems $I^{\pm}_{sh}$ will be canonically equivalent in the extended
phase space of variables $\{P_{\pm},~H_{\pm},~R,~t_{\pm}\}$, if
\begin{equation}\label{b4a}
    d I^{+}_{sh}= d I^{-}_{sh} +dF,
\end{equation}
or
\begin{equation}\label{b4}
     P_{+}dR - H_{+}dt_{+}=P_{-}dR - H_{-}dt_{-}+dF,
\end{equation}
where $F=F(R,t_{+},t_{-} )$ is the generating function of the canonical
transformation $\{P_{+}=P_{+}(P_{-},t_{-}, R),~t_{+}=t_{+}(P_{-},t_{-}, R)
\}$. From (\ref{b4}) we find
\begin{equation}\label{b5}
      H_{+}=-\frac{\partial F}{\partial t_{+}}\, , \quad
      H_{-}= \frac{\partial F}{\partial t_{-}}\, , \quad
      P_{+}= P_{-}-\frac{\partial F}{\partial R}\, .
\end{equation}
Using these relations the constraints (\ref{4.2.14a}) and (\ref{4.2.18})
can be rewritten in the following way
\begin{equation}\label{b5a}
 -\frac{\partial F}{\partial t_{+}}
      =\frac{\partial F}{\partial t_{-}} =E \, , \quad
     \frac{\partial F}{\partial R} =
     P_{-}\left(1 -\frac{f_{-}}{f_{+}} \right)\, .
\end{equation}
From here we find
\begin{equation}\label{b7}
      F=E(t_{-}-t_{+}) + \sigma(R,E)\, ,
\end{equation}
where
\begin{equation}\label{b8a}
      \sigma (R,E)= \frac{1}{c}\int
      \Biggl(\frac{1}{f_{-}}-\frac{1}{f_{+}} \Biggr)
      \sqrt{(E{\pm}U)^2 - m^2 c^4 f_{\pm}}
      \ dR\, .
\end{equation}
The expression under the radical is invariant with respect to the
replacement $f_{\pm}\rightarrow f_{\mp},\, U\rightarrow -U $.

Then differentiating the expression (\ref{b7}) over $E$ one finds the
relation between $t_{+}$ and $t_{-}$
\begin{equation}\label{b9}
      \frac{\partial F}{\partial E} = t_{-}-t_{+}
      + \frac{\partial \sigma}{\partial E} = \alpha\, ,
\end{equation}
where the constant $\alpha $ cab be omitted. Thus, the transformations
\begin{eqnarray}
  \left\{
  \begin{array}{lll}
         t_{+} &=& t_{-} + \frac{\partial \sigma(R,E)}{\partial E}\, , \\
         \\
         P_{+} &=& P_{-} + \frac{\partial \sigma(R,E)}{\partial R}\, , \\
  \end{array}
  \right.
\end{eqnarray}
are the sought canonical transformations of the extended phase space
$\{P_{\pm}, H_{\pm}, R, t_{\pm} \}$ of the system. Herewith, the
corresponding presymplectic form
\begin{equation}\label{b12}
dP_{+}\wedge d R - d E \wedge dt_{+}=dP_{-}\wedge d R - d E \wedge dt_{-}
\end{equation}
is invariant under these transformations.

The difference of the shell actions, which are considered from the point
of view of the exterior and interior observers, up to an additive constant
is $ I^{+}_{sh}(R,t_{+}) - I^{-}_{sh}(R,t_{-})=F(R,t_{+},t_{-})$, where
$F=F(R,t_{+},t_{-})$ is the generating function, is calculated by the
elimination of the energy $E$ from the relations (\ref{b7}) and
(\ref{b9}).

\section{Non-relativistic dust shell}

Let us consider an infinitely thin dust layer in the Euclidean space $
R^{(3)} $ in the form of the closed surface $\Sigma_t$ moving in its own
Newtonian gravitational field $\varphi=\varphi(\vec r)$. The full action
for this configuration has the form
\begin{eqnarray}
I^{~(N)}_{tot} = \int\limits_{t_1}^{t_2}dt~ \biggl\{
      \int\limits_{\Sigma_t}
      \left(
       \frac{1}{2}\sigma \vec v^{\, 2} - \sigma\varphi
      \right)~d^2 f -
      \frac{1}{8\pi\gamma} \int\limits_{D_- \cup D_+}
        (\nabla \varphi)^2~dV
\biggr\}\, .                       \label{1}
\end{eqnarray}
Here  $\Sigma_t \ (t_1 \leq t \leq t_2)$ is a one-parameter family of the
closed surfaces, $D_- $ and $D_+ $ are interior and exterior  regions of
the shell $\Sigma_t $ at a moment $t$, $d^2 f$ is the surface element on
$\Sigma_t $, $d V$ is the volume element in $R^3$, $\vec v$ is the
velocity of particles of the shell, $\nabla$ is the nabla operator,
$\sigma $ is surface mass density of dust on $\Sigma_t $. By virtue of the
mass conservation law the value $ dm\equiv \sigma d^2 f $ conserves along
the stream tube and in case of dust can be considered as a stationary
value under arbitrary variations. Note also, that, we require that the
potential $\varphi$ be continuous, and, together with all its derivatives,
vanish at infinity.

The requirement of extremity of the action $\delta I^{~(N)}_{tot}=0$ with
respect to everywhere continuous variations $\delta  \varphi$, vanishing
on infinity, leads to the Laplace equation
\begin{eqnarray}
\Delta\varphi (\vec r)=0,~~ \vec r \in D_{-} \cup D_{+} \label{3}
\end{eqnarray}
with the boundary conditions for normal derivatives of $\varphi$ on
$\Sigma_t$. The latter, for completeness, will be written out together
with the continuity conditions for $\varphi$ on $\Sigma_t$:
\begin{eqnarray}
     [\varphi]\equiv \varphi _{|_{+}} - \varphi _{|_{-}} = 0\, ,~~~
    \biggl[ \frac{\partial\varphi}{\partial\eta} \biggr]\equiv
\frac{\partial\varphi}{\partial\eta} \biggl|_{+} -~
\frac{\partial\varphi}{\partial\eta} \biggl|_{-} = 4\pi\gamma\sigma\, ,
\quad \vec r \in \Sigma_t \ (t=\mbox{const})\, .           \label{4}
\end{eqnarray}
Here $\partial/\partial\eta =(\vec n\cdot \nabla)$ is the derivative with
respect to the exterior normal  $\vec n$  to $\Sigma_t $, vector $\vec n\
({\vec n}^2=1)$ is directed from $D_{-}$ to $D_{+}$. The solution of the
equations (\ref{3}), (\ref{4}) is the potential of a ``simple layer''
\begin{eqnarray}
\varphi_{\sigma} (\vec r)= -\gamma \int\limits_{\Sigma_t}
          \frac{\sigma(\vec r^{\,\prime})~d^2 f^{\prime}}
         {|\vec r - \vec r^{ \,\prime}|}\, .      \label{5}
\end{eqnarray}

Now we can calculate $ I^{~(N)}_{tot}$ on the solutions of the equations
(\ref{3}), (\ref{4}). Thereby, the potential $\varphi$ is excluded from
the full action (\ref{1}). Note that owing to (\ref{3}) we have $ (\nabla
\varphi)^2=\nabla (\varphi\nabla \varphi) $. This allows to transform the
volume integral of (\ref {1}) into the surface one on the boundaries of
the regions $D_{\pm}$. Taking into account the boundary conditions (\ref
{4}) and an asymptotic behaviour of $\varphi$, we find the reduced action
$I^{~(N)}_{tot}$, as the value of the initial action on the solution
(\ref{5}) of the equations (\ref{3}) and (\ref{4})
\begin{equation}
   {I^{~(N)}_{tot}}_{| \{solutions~ eq.~(\ref{3}),~(\ref{4})\}}
      = I^{(N)}_{sh}+I^{(N)}_{0}.
\end{equation}
Here $I^{~(N)}_{0}$ contains the surface term, which is unessential for
further consideration, and
\begin{eqnarray}
    I^{~(N)}_{sh} = \int \limits_{t_1}^{t_2} L^{(N)}_{sh} ~dt   \label{6}
\end{eqnarray}
is the effective  action for the shell with the Lagrangian
\begin{eqnarray}
L^{(N)}_{sh} = \frac{1}{2}
          \int\limits_{\Sigma_t} \sigma \vec v^{\, 2} df -
 U\, ,                     \label{7}
\end{eqnarray}
where
\begin{eqnarray}
 U = \frac{1}{2}
    \int\limits_{\Sigma_t} \sigma\varphi_{\sigma} ~df
  = -\frac{\gamma}{2} \int\limits_{\Sigma_t}\int\limits_{\Sigma_t}
      \frac{ \sigma(\vec r)~\sigma(\vec r^{\,\prime}) }
           {|\vec r - \vec r^{ \,\prime}|}
      ~df df^{\prime}                            \label{8}
\end{eqnarray}
is the functional of the potential energy of the gravitational self-action
of the shell.

The Lagrangian of the shell in an external gravitational field
$\varphi_0=\varphi_0(\vec r)$ has the form
\begin{eqnarray}
      L^{(N)}_{sh} = \int\limits_{\Sigma_t}
    \left(\frac{1}{2} \sigma \vec v^{\, 2}-\sigma\varphi_0\right)d f
     - U\, .    \label{9}
\end{eqnarray}

Now consider the spherical non-relativistic dust shell. Let $R=R (t)$ be
the radius of the spherical shell at a moment $t$. With respect to the
spherical coordinates $\{ r,\theta,\alpha \}$, we have\ $\sigma =\sigma
(r),\, d^2 f=R(t)\sin\!\theta d\theta d\alpha,\,\vec v^{\, 2}=\dot
R^2=(dR/dt)^2 $. The mass of the shell is $ m=4\pi\sigma R^2 =\mbox{const}
$. Potential of the external field $ \varphi_0 $ on the shell has the
value
\begin{equation}
 \varphi_0 \equiv \varphi_{-} =-\frac{\gamma m_{-}}{R(t)}\, , \label{2.2}
\end{equation}
where $m_{-}$ is the total mass of the interior source. The potential
$\varphi_{\sigma}$ and the self-action energy for the shell $U$ prove to
be equal
\begin{eqnarray}
      \varphi_{\sigma} (r) &=& \left\{
                        \begin{array}{lll}
                        -\gamma m/r,    &r\geq R (t) \\
                        -\gamma m/R (t),&r < R (t)
                        \end{array}
                        \right. \,  ,                     \label{2.3}\\
 U &=& \frac{1}{2}m\varphi_{\sigma} =
             - \frac{\gamma m^2}{2 R(t)}\, .               \label{2.4}
\end{eqnarray}
Replacing the corresponding terms in (\ref{9}) by those of (\ref{2.2}) -
(\ref{2.4}), we obtain the Lagrangian of the spherically-symmetric dust
shell in Newtonian theory of gravity
\begin{eqnarray}
      L^{(N)}_{sh-} = \frac{1}{2} m \dot R^2 + \frac{\gamma mm_{-}}{R}
     - U \, . \label{2.5}
\end{eqnarray}

A distinctive feature of spherical shell is that the two-valued
description of the shell dynamic becomes possible with respect to the
observer's position. From an interior observer's point of view, except for
the force of self-action $-\partial U/\partial R $, the shell is effected
by the external force $F_{-}=-md\varphi_{-}/dr$, which determines an
interior gravitational field. This situation corresponds to the Lagrangian
$L^{(N)}_{sh-}$, therefore the latter can be interpreted as the Lagrangian
describing the non-relativistic shell from an interior observer's point of
view.

An exterior observer $(r>R(t))$ determines the field, judging by the force
$F_{+}=-md\varphi_{+}/dr$ acting on the shell in the field
$\varphi_{+}=\varphi_{-}+\varphi_{\sigma}=-\gamma m_{+} /R(t)$. This field
is generated by the total mass of the system $m_{+}=m_{-}+m$. If, by
making use of this relation, we eliminate $m_{-}$ from $L^{(N)}_{sh-}$ we
obtain the following Lagrangian
\begin{eqnarray}
      L^{(N)}_{sh+} = \frac{1}{2} m \dot R^2 + \frac{\gamma mm_{+}}{R}
                  + U\, . \label{2.6}
\end{eqnarray}
It can be interpreted as the Lagrangian describing the Newtonian shell
from an exterior observer's point of view.

In such a way, transformation from an exterior observer to an interior one
stipulates the discrete transformation  $m_{\pm} \stackrel{~}{\to} m_{\mp}
= m_{\pm} \mp m $. Its can be interpreted as both the gravitational
potential transformation
$\varphi_{\pm}\stackrel{~}{\to}\varphi_{\mp}=\varphi_{\pm}\pm\gamma m/R$
and the change of sign of the self-action potential
$U^{(N)}\stackrel{~}{\to} - U^{(N)}$. The above two-valued description of
spherical shell in the Newtonian theory has a formal character. This
ambiguity, arising when describing spherically-symmetric shell, is a
matter of principle in General Relativity.

Note other feature of the shell, which has non-trivial meaning in General
Relativity. The Lagrangians  $L^{(N)}_{sh\pm}$ completely and closely
determine the motion of boundaries of regions $D_{\pm}$. That is why they
can be thought of as independent systems with their momenta and
Hamiltonians
\begin{eqnarray}
            P_{\pm}=m \dot R_{\pm}\, , \qquad
H_{\pm} = \frac{P^2_{\pm}}{2m} - \frac{\gamma mm_{\pm}}{R_{\pm}}
          \mp U_{\pm} = E_{\pm} \, .
         \label{2.7a}
\end{eqnarray}
Here $E_{\pm}$ are the total energies of these boundaries,
$U^{(N)}_{\pm}=-\gamma m^2 / 2R_{\pm}$ are their potential energies of
self-action, and $R_{\pm}=R_{\pm}(t)$ are the radiuses of the regions'
boundary $D_{\pm}$. The systems (\ref{2.7a}) describe the same shell
provided the regions $D_{\pm}$ have a common boundary $R_{+}(t)=R_{-}(t)
\equiv R(t)$ for all moments $t$. In this case, eliminating the momentum
$P\equiv P_{+}=P_{-}$ from the equations $H_{\pm}=E_{\pm}$ one has
\begin{equation}\label{2.7b}
      E_{+} - E_{-} = \frac{\gamma m}{R} (m_{+}-m-m_{-}) \, .
\end{equation}
Hence it follows an ordinary equality of energies and the additivity of
masses $E_{+}=E_{-},\ m_{+}=m-m_{-}$. In General Relativity, a similar but
not trivial procedure follows from the isometry conditions for the
boundaries of the corresponding four-dimensional regions.

Finally we shall write the corresponding relations for a self-gravitating
shell, when $m_{-}=0$:
\begin{eqnarray}
      L^{(N)}_{sh-} & = & \frac{1}{2} m \dot R^2
     + \frac{\gamma m^2}{2R} \, , \label{2.7}  \\
      P=m \dot R\, , \qquad
      H & = & \frac{P^2}{2m} - \frac{\gamma m^2}{2R} = E \, .\label{2.8}
\end{eqnarray}

\end{document}